\documentclass[twocolumn]{aastex61}
\usepackage{epsfig}

\begin{document}
\title{Fast Extragalactic X-ray Transients From\\ 
         Gamma Ray Bursts Viewed Far Off Axis} 
\author{Shlomo Dado}
\affiliation{Physics Department, Technion, Haifa 32000, Israel}
\author{Arnon Dar}
\affiliation{Physics Department, Technion, Haifa 32000, Israel}

\begin{abstract}

The observed lightcurves and estimated sky rate of 
fast extragalactic x-ray transients (XRTs) discovered 
in archival Chandra data indicate 
that they belong to two distinct XRT populations. The first 
population of relatively short duration pulses, which 
typically last less than few minutes seems to be pulses of x-ray 
flashes (XRFs), which are nearby long duration gamma ray 
bursts viewed from far off axis. The second population of much 
longer pulses, which typically last hours, seems to be the 
early time afterglows of short gamma ray bursts (SGRBs) which 
are beamed away from Earth, as was shown in a previous paper.

\end{abstract}


\section{Introduction}
Two new types of populations of fast extragalactic X-ray 
transients (XRTs) were discovered in the past few years in archival 
images taken before by the Chandra X-ray observatory (CXO). 
Junker et al. (2013) and Glennie et al. (2015) reported the discoveries  
of relatively nearby fast XRT 000519 close to M86 in the Virgo cluster 
at a distance of $\sim 16.5$ Mpc (Mei et al. 2007), and XRT 110103 
in the galaxy cluster ACO 3581 at a distance $\sim 94.9$ Mpc 
(Jhohnstone et al. 2005), respectively, with pulses that lasted less 
than few tens of seconds. XRTs at large cosmological 
distances that lasted more than $10^4$ s were discovered in the 
Chandra Deep Field-South (CDF-S) observations;  CDF-S XT1 associated 
with a faint distant galaxy at unknown redshift, that lasted more 
than a day was discovered by Bauer et al (2017), and CDF-S XT2 
associated with a galaxy at redshift $z=0.738$ that lasted more than 
$2\times 10^4$ s was discovered by Xue et al. (2019).

In a recent publication (Dado and Dar 2019a) we have shown that the 
observed light curve of CDF-S XT2 (Xue et al. 2019) and the 
estimated full sky rate of such XRTs (Yang et al. 2019) indicate 
that they are early time x-ray afterglows of short gamma ray bursts 
(SGRBs), which point away from Earth. Their light curves are 
indistinguishable from those of the early time X-ray afterglows of 
SGRBs which do point to Earth. Their full sky rate is consistent 
with the estimated rate of newly born millisecond pulsars in binary 
neutron stars mergers (Dado and Dar 2019a) whose spin down seems to 
power the early time quasi isotropic afterglow of SGRBs (Dado and 
Dar 2019b).

In this letter we show supportive evidence that the population 
of XRTs which was discovered by Jonker et al. (2013) and by 
Glennie et al. (2015) consists of x-ray flashes (XRFs: Heise et al.
2003; Barraud et al. 2003; Sakamoto et al. 2005), which, 
in the cannonball (CB) model of GRBs, are highly beamed ordinary 
long GRBs (Shaviv and Dar 1995) that are viewed from far-off 
axis (Dar and De Rujula 2000, Dado, Dar and De Rujula 2004). In 
particular, such ordinary GRBs viewed far off-axis were 
predicted by Dar and De R\'ujula (2000, Eqs. 35-40) to have 
much lower luminosity, peak energy and linear polarization, and a
much longer duration, compared to those of ordinary GRBs viewed 
from near axis. As such, they are visible only from relatively 
nearby distances (Dar and De R\'ujula 2000; Dado, Dar and De 
R\'ujula 2004). Moreover, their observed isotopic equivalent 
energy, $E_{iso}$, and peak energy, $E_p$, were predicted (Dar 
and De Rujula 2000, Eq.35) to satisfy the correlation 
$(1+z)E_p\propto [E_{iso}]^{1/3}$ rather than the 
$(1+z)E_p\propto E_{iso}^{1/2}$ correlation, which was later 
discovered empirically by Amati et al. (2002) for ordinary, near 
axis, long duration GRBs (LGRBs).

\section{XRFs properties in the CB model} 
In the CB model, a core collapse supernova of Type Ia
result in a compact remnant and a fast rotating torus 
of non ejected fallen back ejecta. Matter acreting onto the 
compact central object produces a jet of highly 
relativistic plasmoids (CBs) of ordinary matter. 
Glory photons - light emitted/scattered
by the supernova and presupernova ejecta- which undergo
inverse Compton scattering by CB electrons (Shaviv and Dar 1995)
produce a gamma ray pulse  in a narrowly beamed GRB. Although
the detailed properties of these CBs and their emission times 
are not predictable, several correlations and 
various typical properties of
XRF pulses, which result from  a  large viewing angle which
satisfies, $1\gg \theta^2 \gg 1/\gamma^2$
are predictable. They include (Dar and De R\'ujula 2000, Dado, 
Dar and De R\'ujula 2004) the $[E_p,E_{iso}]$ correlation,
\begin{equation}
(1+z)E_p\propto E_{iso}^{1/3}\,,        
\end{equation}
the $[E_p,FWHM]$ "anti" correlation between $E_p$ 
and the full width at half maximum,    
\begin{equation}
E_p\propto 1/FWHM\,,
\end{equation}
the early time large sky projected super luminal velocity,
\begin{equation}
V_{\perp}={\beta\,c\, sin\theta\over 1-\beta\,cos\theta}
          \approx {2\,c\over\theta}\gg c\,,
\end{equation}
the small linear polarization, 
\begin{equation}
P \approx {2\gamma^2 \theta^2\over 1+\gamma^4 \theta^4}\approx 
2/\gamma^2\theta^2\ll 1\,,
\end{equation}
and a typical pulse shape and afterglow light curve.
The pulse shape above a minimal energy $E_m$  has the behavior,
\begin{equation}
{dN_\gamma(E>E_m)\over dt}\!\propto\!{t^2\, 
          exp[-E_m/E_p(0)(1\!-\!t/\sqrt{t^2\!+\!\tau^2})]
          \over (t^2\!+\!\Delta^2)^2}\,
\end{equation}

For XRFs with $\tau\! \gg \!\Delta$, the exponential factor on the 
right hand side of eq.(5) can be neglected, which yields  
a full width at half maximum $FWHM\!\approx 2\,\Delta$, a rise time
$RT\!\approx \!0.59\,\Delta$ from half to peak value at $t=\Delta$,
and a decay time $DT\!\approx \!1.41\,\Delta$ from peak to half peak. 

\section{Comparison with observations}

To test whether the x-ray transients XRT 000519 (Jonker et al. 2013) and 
XRT 110103 (Glennie et al. 2015) and CDF-S XT1 (Bauer et al. 2017) in the 
Chandra archival data could have been XRFs, i.e., LGRBs viewed far 
off-axis, we have fitted the 0.3-7 keV light curves of their pulses with 
Eq.(5). These fits with quite a satisfactory $\chi^2/dof$ are shown in 
Figures 1,2.

\begin{figure}[]
\centering  
\vbox{
\epsfig{file=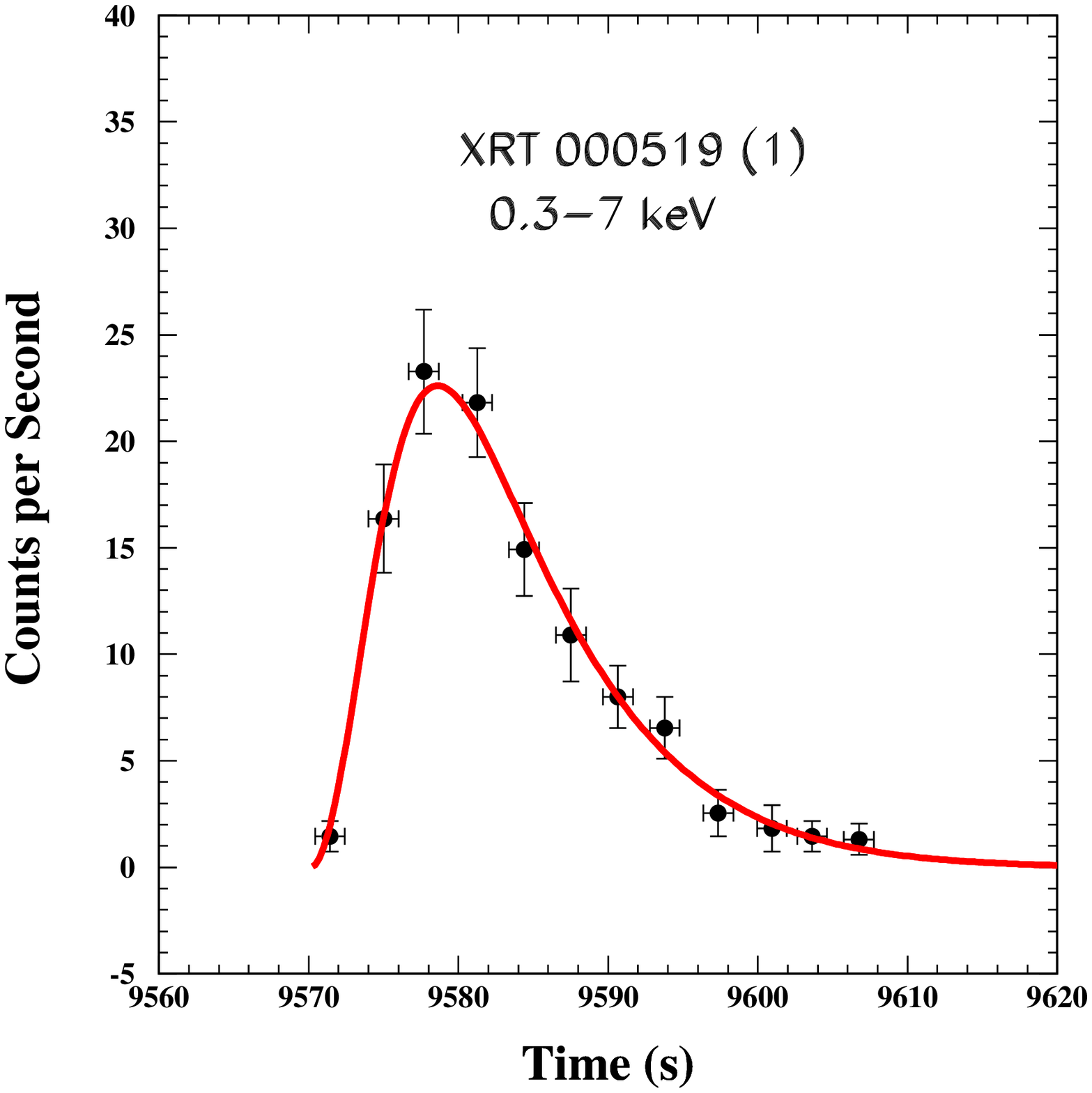,width=9.cm,height=10.cm}
\epsfig{file=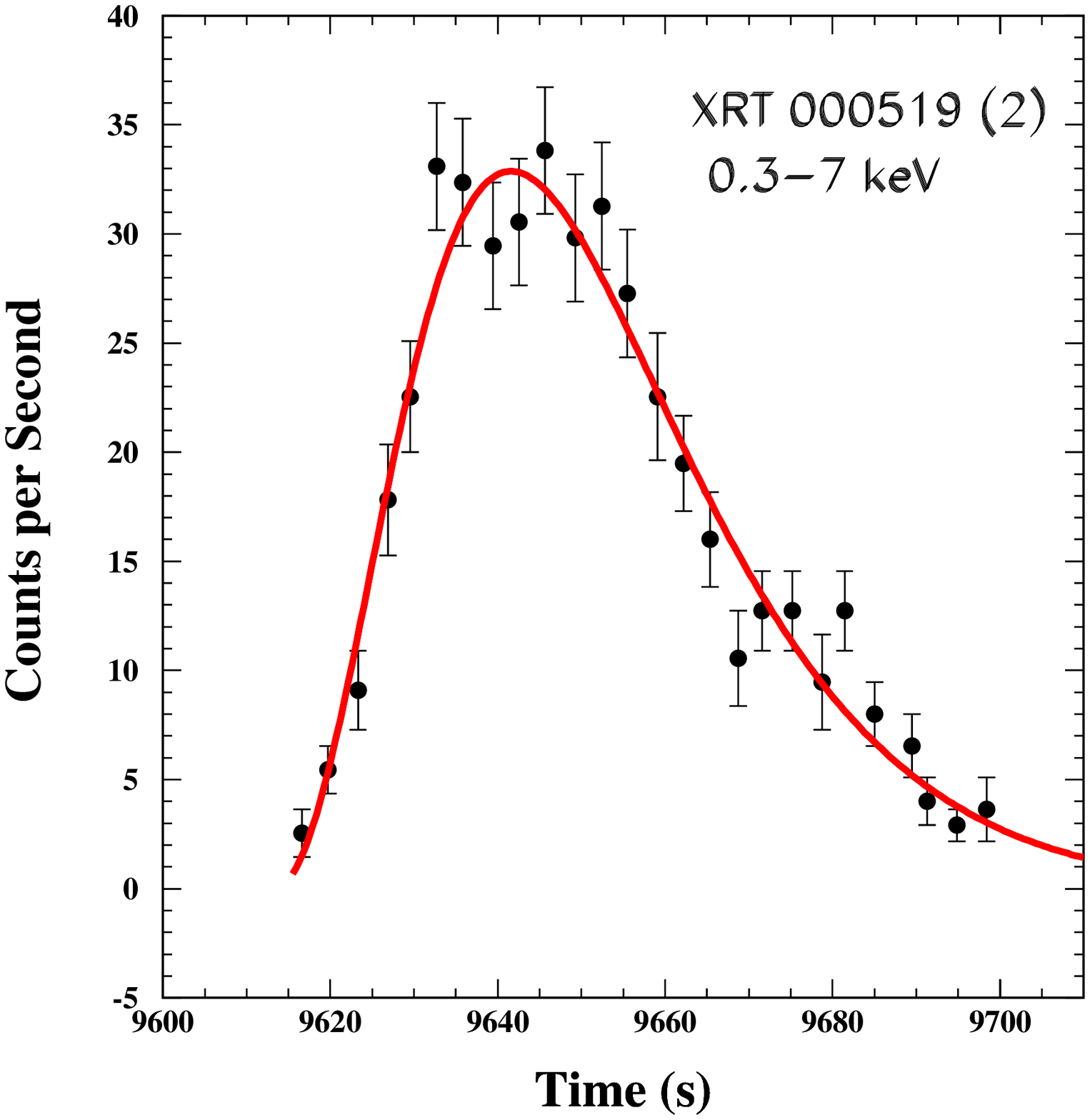,width=9.cm,height=10.cm}
}
\caption{{\bf Top:} Comparison between the observed
pulse shape of the first pulse of XRT 000519 (Jonke  et al. 2013) in the 0.3-7 keV 
x-ray band and  eq.(5) with
the best fit parameters listed in Table I,
which yield a $\chi^2/df=0.39$.
{\bf Bottom:} Comparison between the observed
pulse shape of the second pulse of 000519 
(Jonker et al. 2013) in the 0.3-7 keV
x-ray band and  eq.(5) with
the best fit parameters listed in Table I,
which  yield $\chi^2/df=1.28$.}
\end{figure}

\begin{figure}[]
\centering
\vbox{
\epsfig{file=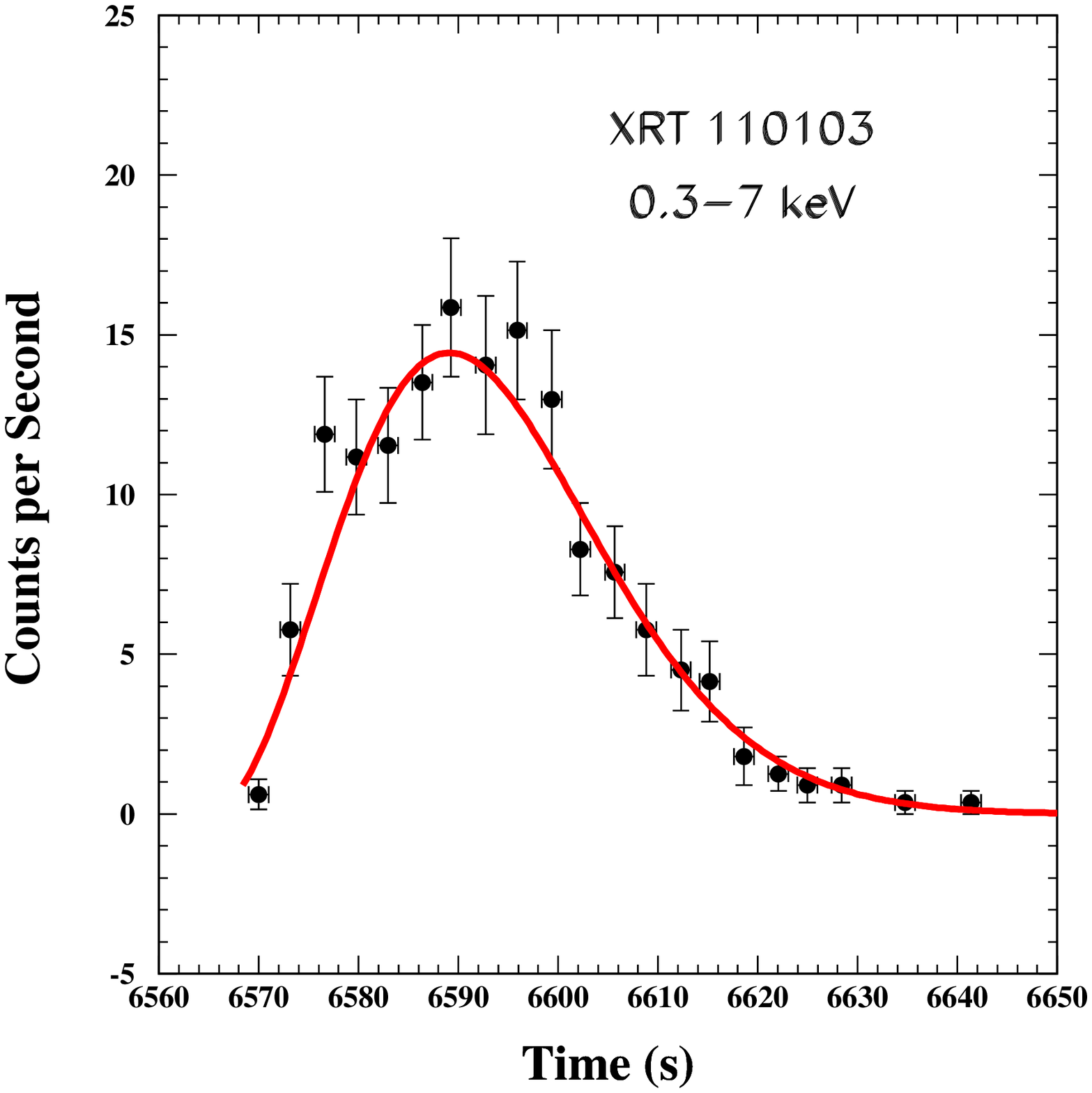,width=9.cm,height=10.cm}
\epsfig{file=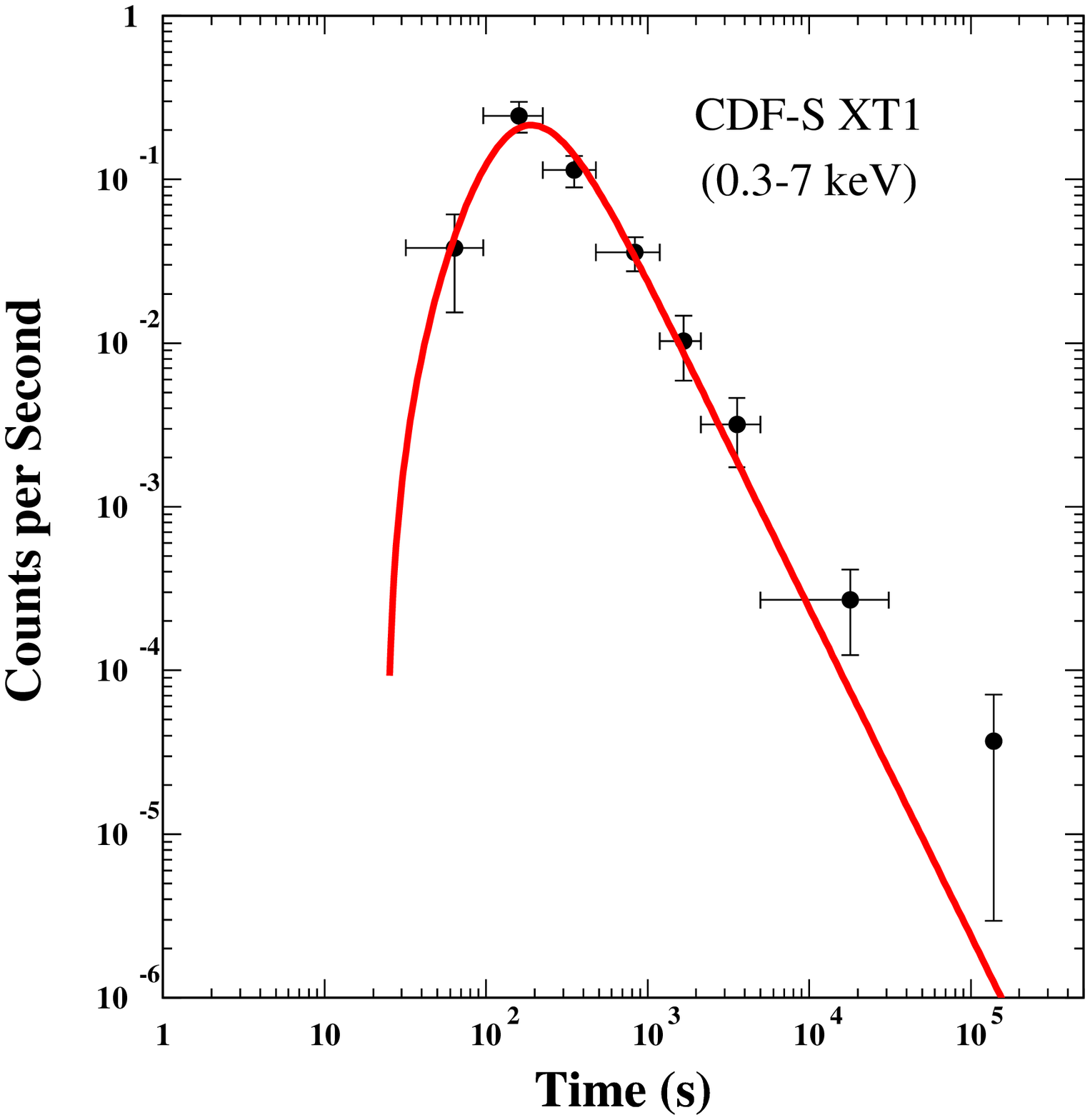,width=9.cm,height=10.cm}
}
\caption{{\bf Top:} Comparison between the observed
pulse shape of XRT 110103 (Glennie et al. 2015) in the 0.3-7 keV
x-ray band and  eq.(5) with
the best fit parameters listed in Table I,
which yield a $\chi^2/df=1.20$.\\
{\bf Bottom:} Comparison between the observed
pulse shape of XDF-S XT1 (Bauer et al. 2017) in the 0.3-7 keV
x-ray band and  eq.(5) with  
the best fit parameters listed in Table I, 
which  yield $\chi^2/df=1.13$.}
\end{figure}

To test further whether the x-ray transients XRT 000519 (Jonker et 
al. 2013) and XRT 110103 (Glennie et al. 2015) discovered in the 
Chandra archival data could have been XRFs we have plotted the best 
fit CB model correlation $(1+z)\,E_p\propto E_{iso}^{1/3}$ (solid 
line) obtained for low luminosity long GRBs, where we included XRT 
000519. As shown in Figure 3 the values $E_p\approx 1.5\pm 0.5$ keV 
and $E_{iso}\approx (4\pm 2)\times 10^{44}$ erg, reported by
Bauer et al. (2017) satisfy well the CB model $[E_p,E_{iso}]$ 
correlation obeyed by XRFs.

\begin{figure}[]
\centering
\epsfig{file=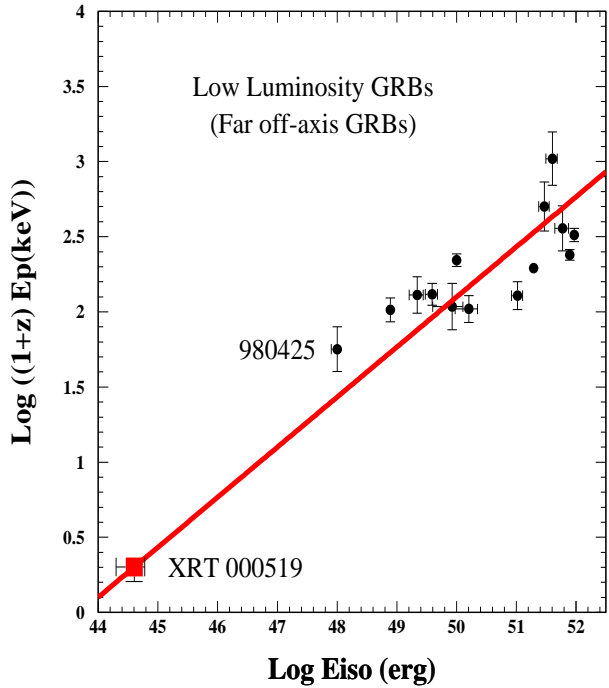,width=9.cm,height=10.cm}
\caption{The $[E_p,E_{iso}]$ correlation
in LGRBs viewed far off axis (which include the so called 
low-luminosity LGRBs and XRFs) and  XRT 000519, which is
indicated by a full (red) square.
The line is the CB model predicted correlation 
for LGRBs viewed far off axis,  given by Eq.(5).}
\end{figure}

\section{Conclusions}

The observed lightcurves, sky rate and distances of the population of 
nearby extragalactic fast x-ray transients with a 
duration less than few minutes  
discovered in archival Chandra data (Jonker et al. 2013, 
Glennie et al. 2015) are consistent with being ordinary GRB pulses of 
ordinary long duration GRBs viewed from far off axis. This population is 
different from the population of much longer (hours) duration of XRTs at 
large cosmological distances discovered also in archival Chandra data 
(Bauer et al. 2017, Xue et al. 2019) which seems to be the early time 
isotropic x-ray afterglow of SGRBs which beamed away from Earth (Xue et 
al 2019, Dado and Dar 2019a).

\begin{table*}
\caption{Best fit parametess of XRT pulse shape}
\label{table1}
\centering
\begin{tabular}{l l l l l l l}
\hline
~~~ Pulse~~~ & start t [s]& $\Delta$ [s] & $\tau $ [s] & $E_p(0)$ [keV]& 
$\chi^2/dof $ \\
\hline
000519 P1 & 9570.10 & ~9.65 & 17.50 & 1.37 & 0.39 \\
000519 P2 & 9613.28 & 35.54 & 28.58 & 2.71 & 1.28 \\
110103 P1 & 6564.90 & 45.03 & 23.16 & 1.22 & 1.20 \\
CDF-S XT1 & 23.48   &167.06 & $>>\Delta$ & & 1.13 \\
\hline
\end{tabular}
\end{table*}

\end{document}